\documentclass[twocolumn,showpacs,preprintnumbers,amsmath,amssymb]{revtex4}

\usepackage{graphicx}
\usepackage{dcolumn}
\usepackage{bm}

\begin{document}

\footnote{$^{*}$Supported by the National Natural Science
Foundation of China under Grant Nos 10373005, 10673002 and
10778616.} \footnote{$^{**}$To whom correspondence should be
addressed. Email: zhangbo@hebtu.edu.cn}


\title{Neutron-Capture Elements in the Double-Enhanced
Star HE 1305-0007: a New s- and r-Process Paradigm$^{*}$}

\author{CUI Wen-Yuan(´ÞÎÄÔª)$^{1,2,3}$, CUI Dong-Nuan(´Þ¶¬Å¯)$^{1}$,
DU Yun-Shuang(¶ÅÔÆ˪)$^{1}$, ZHANG Bo(ÕŲ¨)$^{1,2}$}

\affiliation{$^{1}$Department of Physics, Hebei Normal University,
Shijiazhuang 050016\\
\ $^{2}$National Astronomical Observatories,
Chinese Academy of Sciences,  Beijing 100012 \\
\ $^{3}$Graduate School of the Chinese Academy of Sciences,
Beijing 100049}

(Received 12 January 2007)

\begin{abstract}
The star HE 1305-0007 is a metal-poor double-enhanced star with
metallicity [Fe/H] $=-2.0$, which is just at the upper limit of
the metallicity for the observed double-enhanced stars. Using a
parametric model, we find that almost all s-elements were made in
a single neutron exposure. This star should be a member of a
post-common-envelope binary. After the s-process material has
experienced only one neutron exposure in the nucleosynthesis
region and is dredged-up to its envelope, the AGB evolution is
terminated by the onset of common-envelope evolution. Based on the
high radial-velocity of HE 1305-0007, we speculate that the star
could be a runaway star from a binary system, in which the AIC
event has occurred and produced the r-process elements.
\end{abstract}


\maketitle

The discovery that several stars show enhancements of both
r-process and s-process elements (s+r stars hereafter)$^{[1,2]}$
is puzzling, as they require pollution from both an AGB star and a
supernova. In 2003, Qian and Wasserburg$^{[3]}$ proposed a theory,
i.e. accretion-induced collapse(AIC), for the possible creation of
s+r-process stars. Another possible s+r scenario is that the AGB
star transfers s-rich matter to the observed star but not suffer
from a large mass loss and at the end of the AGB phase, the
degenerate core of low-metallicity, high-mass AGB star may reach
the Chandresekhar mass, leading to type-1.5 supernova.$^{[4]}$
Because the initial-final-mass relation flats at higher
metallicity,$^{[4]}$ the degenerate cores of high-metallicity AGB
stars are smaller than those of the low-metallicity stars, the
formation of AIC or SN1.5 is more difficult in the
high-metallicity binary system, which can explain the upper limit
of the metallicity ([Fe/H] $<-2.0$) for the observed r+s
stars.$^{[5]}$ Recently, Barbuy {\it et al.}$^{[6]}$ and Wanajo
{\it et al.}$^{[7]}$ suggested massive AGB stars
($M=8\sim12$M$_\odot$) to be the origin of these double
enhancements. Such a large mass AGB star could possibly provide
the observed enhancement of s-process elements in the first phase,
and explode or collapse providing the r-process elements. However,
the modeling of the evolution of such a large mass metal-poor star
is a difficult task, an amount of the s-process material is
produced and its abundance distribution is still
uncertain.$^{[7]}$

The generally favoured s-process model till now is associated with
the partial mixing of protons (PMP hereafter) into the radiative
C-rich layers during thermal pulses.$^{[8-11]}$ PMP activates the
chain of reactions $^{12}$C(p,
$\gamma$)$^{13}$N($\beta$)$^{13}$C($\alpha$, n)$^{16}$O, which
likely occurs in a narrow mass region of the He intershell (i.e.
$^{13}$C-pocket) during the interpulse phases of an AGB star. The
nucleosynthesis of neutron-capture elements in the carbon-enhanced
metal-poor stars (CEMP stars hereafter)$^{[12]}$ can be
investigated by abundance studies of s-rich or r-rich stars. In
2006, Goswami {\it et al.}$^{[13]}$ analysed the spectra of the
s-and r-rich metal-poor star HE 1305-0007, and concluded that the
observed abundances could not be well fit by a scaled solar system
r-process pattern nor by the s-process pattern of an AGB model.
This star shows that the enhancements of the neutron-capture
elements Sr and Y are much lower than the enhancement of Ba and
the abundances ratio [Pb/Ba] is only about 0.05. Because of the Na
overabundance, which is believed to be formed through deep
CNO-burning, Goswami {\it et al.}$^{[13]}$ have also speculated
that this star should be polluted by a massive AGB star. Clearly,
the restudy of elemental abundances in this object is still very
important for well understanding the nucleosynthesis of
neutron-capture elements in metal-poor stars.

The chemical abundance distributions of the very metal-poor
double-enhanced stars are excellent information to set new
constraints on models of neutron-capture processes at low
metallicity. The metallicity of HE 1305-0007 is [Fe/H] $=-2.0$,
which is just at the upper limit of the metallicity for the
observed double-enhanced stars. There have been many theoretical
studies of s-process nucleosynthesis in low-mass AGB stars.
Unfortunately, the precise mechanism for chemical mixing of
protons from the hydrogen-rich envelope into the $^{12}$C-rich
layer to form a $^{13}$C-pocket is still unknown.$^{[14]}$ It is
interesting to adopt the parametric model for metal-poor stars
presented by Aoki {\it et al.}$^{[15]}$ and developed by Zhang
{\it et al.}$^{[5]}$ to study the physical conditions which could
reproduce the observed abundance pattern found in this star. In
this Letter, we investigate the characteristics of the
nucleosynthesis pathway that produces the special abundance ratios
of s- and r-rich object HE 1305-0007 using the s-process
parametric model.$^{[5]}$ The calculated results are presented. We
also discuss the characteristics of the s-process nucleosynthesis
at low metallicity.

We explored the origin of the neutron-capture elements in HE
1305-0007 by comparing the observed abundances with predicted s-
and r-process contribution. For this purpose, we adopt the
parametric model for metal-poor stars presented by Zhang {\it et
al.}$^{[5]}$ The $i$th element abundance in the envelope of the
star can be calculated by
\begin{equation}
N_{i}(Z)=C_{s}N_{i,s}+C_rN_{i,r}10^{[Fe/H]},
\end{equation}
where $Z$ is the metallicity of the star, $N_{i,s}$ is the
abundance of the \textit{i}-th element produced by the s-process
in the AGB star and $N_{i,r}$ is the abundance of the $i$th
element produced by the r-process (per Si $=10^6$ at $Z=Z_\odot$),
$C_s$ and $C_r$ are the component coefficients that correspond to
contributions from the s-process and r-process respectively.

There are four parameters in the parametric model of s- and r-rich
stars. They are the neutron exposure per thermal pulse
$\Delta\tau$, the overlap factor $r$, the component coefficient of
the s-process $C_{s}$ and the component coefficient of the
r-process $C_{r}$. The adopted initial abundances of seed nuclei
lighter than the iron peak elements were taken to be the
solar-system abundances, scaled to the value of [Fe/H] of the
star. Because the neutron-capture-element component of the
interstellar gas to form very mental-deficient stars is expected
to consist of mostly pure r-process elements, for the other
heavier nuclei we use the r-process abundances of the solar
system,$^{[16]}$ normalized to the value of [Fe/H]. The abundances
of r-process nuclei in Eq. (1) are taken to be the solar-system
r-process abundances$^{[16]}$ for the elements heavier than Ba,
for the other lighter nuclei we use solar-system r-process
abundances multiplied by a factor of 0.4266.$^{[5,17]}$ Using the
observed data in the sample star HE 1305-0007, the parameters in
the model can be obtained from the parametric approach.

Figure 1 shows our calculated best-fit result. For this star, the
curves produced by the model are consistent with the observed
abundances within the error limits. The agreement of the model
results with the observations provides strong support to the
validity of the parametric model. In the AGB model, the overlap
factor $r$ is a fundamental parameter. In 1998, Gallino {\it et
al.}$^{[8]}$ (G98 hereafter) have found an overlap factor of $r
\simeq 0.4-0.7$ in their standard evolution model of low-mass
($1.5-3.0$M$_{\odot}$) AGB stars at solar metallicity. The overlap
factor calculated for other s-enhanced metal-poor stars lies
between 0.1 and 0.81.$^{[5]}$ The overlap factor deduced for HE
1305-0007 is about $r=0^{+0.17}_{-0.00}$, which is much smaller
than the range presented by G98. This just implies that iron seeds
could experience only one neutron exposure in the nucleosynthesis
region.$^{[18]}$
\begin{figure}
\includegraphics[width=0.5\textwidth,height=0.25\textheight]{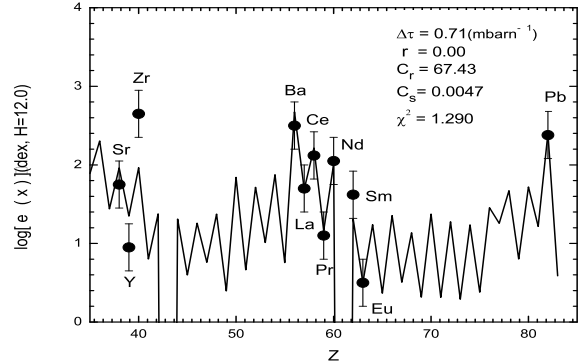}
\caption{\label{fig:epsart} Best fit to observational result of HE
1305-0007. The black circles with appropriate error bars denote
the observed element abundances, the solid line represents
predictions from s-process calculations considering r-process
contribution (taken from Ref.\,[13]).}
\end{figure}

For the third dredge-up and the AGB model, several important
properties depend primarily on the core mass.$^{[19-21]}$ In the
core-mass range $0.6\leq M_{c}\leq1.36$, an analytical formula for
the AGB stars was given by Iben$^{[19]}$ showing that the overlap
factor increases with decreasing core mass. Combing the formula
and the initial-final mass relations,$^{[4]}$ Cui and Zhang
$^{[22]}$ obtained the overlap factor as a function of the initial
mass and metallicity. In an evolution model of AGB stars, a small
$r$ may be realized if the third dredge-up is deep enough for the
s-processed material to be diluted by extensive admixture of
unprocessed material. Karakas$^{[21]}$ and Herwig$^{[23,24]}$ have
found that the third dredge-up is more efficient for the AGB stars
with larger core masses, confirming the low values of $r$ obtained
by Iben$^{[19]}$ in these cases. In AGB stars with initial mass in
the range $M=1.0-4.0$M$_\odot$, the core mass $M_c$ lies between
0.6 and 1.2M$_\odot$ at [Fe/H] $=-2.0$. According to the formula
presented by Iben,$^{[19]}$ the corresponding values of $r$ would
range between 0.76 and 0.26. Obviously, the overlap factor of HE
1305-0007 is smaller than this range.
\begin{figure}
\includegraphics[width=0.5\textwidth,height=0.32\textheight]{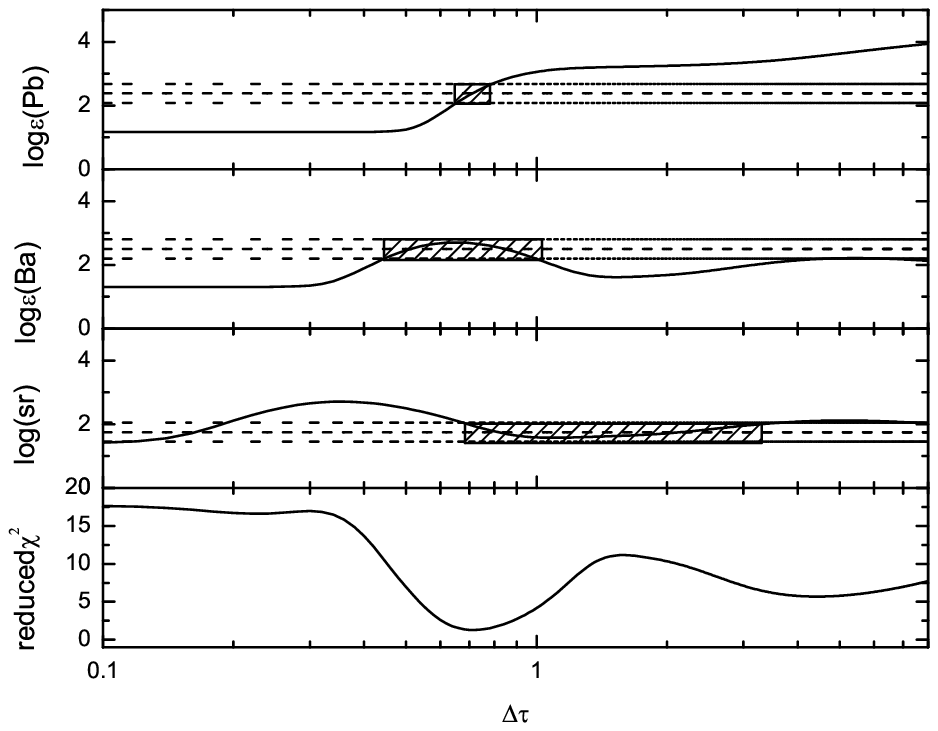}
\caption{\label{fig:epsart} Best fit to observational result of
metal-deficient star HE 1305-0007 shows the calculated abundances
log$\varepsilon$(Pb), log$\varepsilon$(Ba) and
log$\varepsilon$(Sr) and reduced $\chi^2$ (bottom)as a function of
the neutron exposure $\Delta\tau$ in a model with $C_r=67.4$,
$C_s=0.0047$ and $r=0$. These are compared with the observed
abundances of HE 1305-0007.}

\includegraphics[width=0.5\textwidth,height=0.32\textheight]{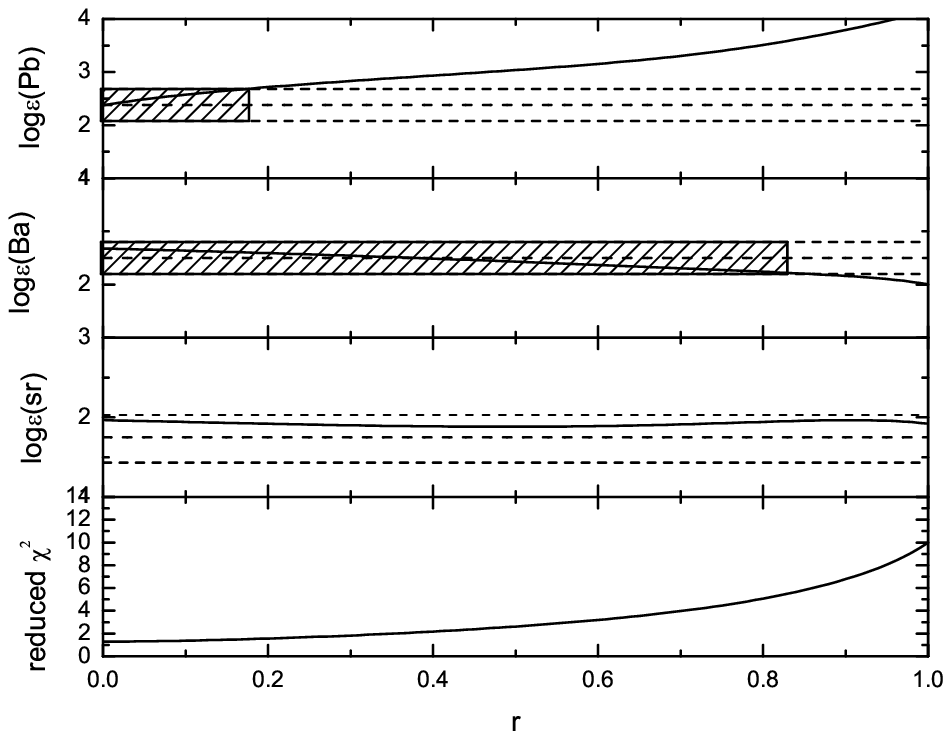}
\caption{\label{fig:epsart} The same as those in Fig. 2 but as a
function of the overlap factor $r$ in a model with
$\Delta\tau=0.71$.}
\end{figure}

We have extensively explored the convergence of the abundance
distribution of s-process elements through recurrent neutron
exposures. All elements, including Pb, were found to be made in
the first neutron exposure. This is consistent with the small
overlap factor $r\simeq0$ deduced in our best-fit model. Thus the
possibility that the s-process material has experienced only one
neutron exposure in the nucleosynthesis region is existent.

In 2000, Fujimoto, Ikeda and Iben$^{[25]}$ have proposed a
scenario for the extra-metal-poor AGB stars with [Fe/H]$<-2.5$ in
which the convective shell triggered by the thermal runaway
develops inside the helium layer. Once this occurs,$^{12}$C
captures proton to synthesize$^{13}$C and other neutron-source
nuclei. The thermal runaway continues to heat material in the
thermal pulse so that neutrons produced by the $^{22}$Ne($\alpha$,
n)$^{25}$Mg reaction as well as the $^{13}$C($\alpha$, n)$^{16}$O
reaction may contribute. In this case, only one episode of proton
mixing into He intershell layer occurs in metal-poor
stars.$^{[25,15,45]}$ After the first two pulses no more proton
mixing occurs although the third dredge-up events continue to
repeat, so the abundances of the s-rich metal-poor stars can be
characterized by only one neutron exposure. Obviously, the
metallicity of HE 1305-0007 is higher than the range of
metallicity for this scenario.

One major goal of this work is to explore the characteristics of
the binary system that HE 1305-0007 origin belongs to. The
enhancement of the neutron-capture elements Ba and Pb suggests
that in a binary system a mass-transfer episode from a former AGB
star took place. The radial-velocity measurement indicates that HE
1305-0007 is a high-velocity star, with a radial-velocity of 217.8
km s$^{-1}$. From the high velocity of HE 1305-0007, we could
speculate that the star could be a runaway star from a binary
system, which has experienced the AIC event. The strong
overabundance of r-process elements for HE 1305-0007 ($C_r=67.4$)
should be a significant evidence for the AIC scenario. In this
case, the orbital separation must be small enough to allow for
capture of a sufficient amount of material to create the formation
of this event. Assuming that HE 1305-0007 is formed in a binary
system, the AGB connection strongly suggests that this star is a
member of a post-common-envelope binary. This must be the case if
the overabundances of s-process elements are attributed to
mass-transfer from an AGB star. We can only speculate about the
effects of common-envelope phase on the nuclear signatures in a
metal-poor star that was formed from this mechanism. One case
could involve several thermal pulses with dredge-up causing the
observed abundance distribution corresponding to larger overlap
factor. However, after the s-process material has experienced only
one neutron exposure in the nucleosynthesis region and is
dredged-up to its envelope, the AGB evolution is terminated by the
onset of common-envelope evolution. This could explain the
characteristic of single neutron exposure in this star. In
addition, based on the Na overabundance, Goswami {\it et
al.}$^{[13]}$ have speculated that HE 1305-0007 should be polluted
by a massive AGB star, which has a large core-mass and favours the
formation of AIC. Clearly, a detailed theoretical investigation of
this scenario is highly desirable.

The neutron exposure per pulse, $\Delta\tau$, is another
fundamental parameter in the AGB model. In 2006, Zhang {\it et
al.}$^{[5]}$ have deduced the neutron exposure per pulse for other
s-enhanced metal-poor stars which lies between 0.45 and 0.88
mbarn$^{-1}$. The neutron exposure deduced for HE 1305-0007 is
about $\Delta\tau=0.71^{+0.06}_{-0.04}$ mbarn$^{-1}$. Figures 2
and 3 show the calculated abundances log$\varepsilon$(Pb),
log$\varepsilon$(Ba) and log$\varepsilon$(Sr) as versus the
neutron exposure $\Delta\tau$ in a model with $C_r=67.4$,
$C_s=0.0047$ and $r=0$ and versus overlap $r$ with
$\Delta\tau=0.71$ mbarn$^{-1}$, respectively. These are compared
with the observed abundances of HE 1305-0007. There is only one
region in Fig.\,2, $\Delta\tau=0.71^{+0.06}_{-0.04}$ mbarn$^{-1}$,
in which all the observed ratios of three representative elements
can be accounted for within the error limits. The bottom panel in
Fig. 2 displays the reduced $\chi^2$ value calculated in our model
with all detected elemental abundances being taken into account
and there is a minimum, with $\chi^2=1.290$, at $\Delta\tau=0.71$
mbarn$^{-1}$. From Fig. 3, we find that the abundances
log$\varepsilon$(Pb), log$\varepsilon$(Ba) and
log$\varepsilon$(Sr)are insensitive to the overlap factor $r$ in a
wider range, $0\leq r\leq0.17$. The uncertainties of the
parameters for the star HE 1305-0007 are similar to those for
metal-poor stars LP 625-44 and LP 706-7 obtained by Aoki {\it et
al.}$^{[15]}$

In addition, it is worth further commenting on the behaviour of
log$\varepsilon$(Sr), log$\varepsilon$(Ba) and
log$\varepsilon$(Pb) as a function of the neutron exposure
$\Delta\tau$ seen in Fig. 2. The nonlinear trends displayed in the
plot reveal the complex dependence on the neutron exposure. The
trends can be illustrated as follows. Starting from low neutron
exposure and moving toward higher neutron exposure values, they
show how the Sr peak elements are preferentially produced at
nearly $\Delta\tau$$\sim$ 0.4mbarn$^{-1}$. At larger neutron
exposure (e.g., $\Delta\tau$$\sim$ 0.7mbarn$^{-1}$), the Ba-peak
elements become dominant. In fact, the higher neutron exposure
favors large amounts of production of the heavier elements such as
Ba, La, etc. and less Sr, Y, etc.,$^{[22]}$ which is the reason of
the abundance pattern of the s-process elements in HE 1305-0007,
i.e. the enhancements of the neutron-capture elements Sr and Y are
much lower than the enhancement of Ba and the abundances ratio
[Pb/Ba] is only about 0.05. Then a higher value of
log$\varepsilon$(Pb)$\sim4$ follows at $\Delta\tau=1.5$
mbarn$^{-1}$. In this case, the s-process flow extends beyond the
Sr-peak and Ba-peak nuclei to cause an accumulation at $^{208}$Pb.
Clearly, log$\varepsilon$(Pb) is very sensitive to the neutron
exposure.

The r- and s-process component coefficients of HE 1305-0007 are
about 67.4 and 0.0047, which implies that this star belongs to s+r
stars. Recently, Zhang {\it et al.}$^{[5]}$ have calculated 12 s+r
stars with $0.0005\leq C_{s}\leq0.0060$. The s-process component
coefficient of HE 1305-0007 lies in this range. The Ba and Eu
abundances are most useful for unraveling the sites and nuclear
parameters associated with the s- and r-process corresponding to
those in extremely metal-poor stars, polluted by material with a
few times of nucleosynthesis processing. In the Sun, the elemental
abundances of Ba and Eu consist of significantly different
combinations of s- and r-process isotope contributions, with s:r
ratios for Ba and Eu of 81:19 and 6:94, respectively.$^{[16]}$
From Eq.\,(1), we can obtain the s:r ratios for Ba and Eu are
95.7:4.3 and 30.1:69.9, which are obviously larger than the ratios
in the solar system. From Fig.\,1 we find that our model cannot
explain the larger errors of some neutron-capture elements, such
as Y and Zr in HE 1305-0007. This implies that our understanding
of the true nature of s-process or r-process is incomplete for at
least some of these elements.$^{[27]}$

In conclusion, the star HE 1305-0007 is an s+r star with
metallicity [Fe/H] $=-2.0$, which is just at the upper limit of
the metallicity for the observed double-enhanced stars.
Theoretical predictions for abundances starting with Sr fit well
the observed data for the sample star, providing an estimation for
neutron exposure occurred in AGB star. The calculated results
indicated that almost all s-elements were made in the first
neutron exposure. Once this happens, after only one time
dredge-up, the observed abundance profile of the s-rich stars may
be reproduced in a single neutron exposure. From the high
radial-velocity of HE 1305-0007, we speculate that the star could
be a runaway star from a binary system, which has experienced the
AIC event. The r-process elements in HE 1305-0007 ($Cr=67.4$)
should come from  the AIC event. Because the orbital separation
must be small enough to allow for capture of a sufficient amount
of material to create the formation of AIC, this star should be a
member of a post-common-envelope binary. After the s-process
material has experienced only one neutron exposure in the
nucleosynthesis region and is dredged-up to its envelope, the AGB
evolution is terminated by the onset of common-envelope evolution.
Clearly, such an idea requires a more detailed high-resolution
study and long-term radial-velocity monitoring in order to reach a
definitive conclusion. More in-depth theoretical and observational
studies of this scenario is highly desirable.

\vspace{0.5cm} \hskip 7pt
 {\bf References}

$[1]$ Hill V et al 2000 {\it Astron. Astrophys.} {\bf 353} 557

$[2]$ Cohen J G et al 2003 {\it Astrophys. J.} {\bf 588} 1082

$[3]$ Qian Y Z and Wasserburg G J 2003 {\it Astrophys. J.} {\bf
588} 1099

$[4]$ Zijlstra A A 2004 {\it Mon. Not. R. Astron. Soc.} 348, L23

$[5]$ Zhang B, Ma K and Zhou G D 2006 {\it Astrophys. J.} {\bf
642} 1075

$[6]$ Barbuy B et al 2005 {\it Astron. Astrophys.} {\bf 429} 1031

$[7]$ Wanajo S et al 2005 {\it Astrophys. J.} {\bf 636} 842

$[8]$ Gallino R et al 1998 {\it Astrophys. J.} {\bf 497} 388

$[9]$ Gallino R et al 2003 Nucl. Phys. A {\bf 718} 181

$[10]$ Straniero O et al 1995 {\it Astrophys. J.} {\bf 440} L85

$[11]$ Straniero O, Gallino R and Cristallo S 2006 Nucl. Phys. A
{\bf 777} 311

$[12]$ Cohen J G et al 2005 {\it Astrophys. J.} {\bf 633} L109

$[13]$ Aruna Goswami et al 2006 {\it Mon. Not. R. Astron. Soc.}
{\bf 372} 343

$[14]$ Busso M et al 2001 {\it Astrophys. J.} {\bf 557} 802

$[15]$ Aoki W et al 2001 {\it Astrophys. J.} {\bf 561} 346

$[16]$ Arlandini C et al 1999 {\it Astrophys. J.} {\bf 525} 886

$[17]$ Cui W Y et al 2007 {\it Astrophys. J.} {\bf 657} 1037

$[18]$ Ma K, Cui W Y and Zhang B 2007 {\it Mon. Not. R. Astron.
Soc.} {\bf 375} 1418

$[19]$ Iben I Jr 1977 {\it Astrophys. J.} {\bf 217} 788

$[20]$ Groenewegen M A T and de Jong T 1993 {\it Astron.
Astrophys.} {\bf 267} 410

$[21]$ Karakas A I, Lattanzio J C and Pols O R 2002 PASA {\bf 19}
515

$[22]$ Cui W Y and Zhang B 2006 {\it Mon. Not. R. Astron. Soc.}
{\bf 368} 305

$[23]$ Herwig F 2000 {\it Astron. Astrophys.} {\bf 360} 952

$[24]$ Herwig F 2004 {\it Astrophys. J.} {\bf 605} 425

$[25]$ Fujimoto M Y, Ikeda Y and Iben I Jr 2000 {\it Astrophys.
J.} {\bf 529} L25

$[26]$ Iwamoto N et al 2003 Nucl. Phys. A {\bf 718} 193

$[27]$ Travaglio C et al 2004 {\it Astrophys. J.} {\bf 601} 864

\end{document}